\documentclass[runningheads]{llncs}
\usepackage{graphicx}
\usepackage{multirow}
\usepackage{array}
\usepackage{xcolor}
\usepackage{float}
\begin{document}

\title{Users' Perception of Search Engine Biases and Satisfaction}
\author{Bin Han \and
Chirag Shah \and
Daniel Saelid}
\institute{University of Washington, Seattle, Washington, USA}
\maketitle           

\begin{abstract}
Search engines could consistently favor certain values over the others, which is considered as biased due to the built-in infrastructures. Many studies have been dedicated to detect, control, and mitigate the impacts of the biases from the perspectives of search engines themselves. In our study, we take the perspective from end-users to analyze their perceptions of search engine biases and their satisfaction when the biases are regulated. In the study, we paired a real search page from search engine Bing with a synthesized page that has more diversities in the results (i.e. less biased). Both pages show the top-10 search items given search queries and we asked participants which one do they prefer and why do they prefer the one selected. Statistical analyses revealed that overall, participants prefer the original Bing pages and the locations where the diversities are introduced are also associated with users' preferences. We found out that users prefer results that are more consistent and relevant to the search queries. Introducing diversities undermines the relevance of the search results and impairs users' satisfaction to some degree. Additionally, we confirmed that users tend to pay more attention to the top portion of the results than the bottom ones.

\end{abstract}

\section{Introduction}

Search engines often present results that are biased toward one subtopic, view, or perspective due to the way they compute relevance and measure user satisfaction. Among various types of search engine biases, one describes the case where the search engines embed features that favor certain values over the others \cite{sep-ethics-search,Eric-bias}. Many studies have attempted to detect, measure and mitigate the impacts from search engines biases. All those works aimed to address the issues from the source of the biases --- search engines themselves. 

In the previous study we conducted [under review], we took a different path to inspect the problem from the aspect of end-users. We paired a real search page and a synthesized page (more varieties in the search results, thus less biased) and asked participants which one do they prefer. The results showed no significant differences between the ratios of selecting two pages. However, what remained unknown to us is that why did participants select the ones they prefer? What are the reasonings underneath their preferences? Therefore, we revisited this study and improved our survey design catering to our goals (more details in Section \ref{method}). We would like to evaluate users' perceptions of the biases, thus hoping to reveal the reasoning of their preferences. Additionally, we are interested in studying the effects on users' satisfactions when the biases are regulated.

\section{Background}\label{literature}

Several prior studies have attempted to disclose and regulate biases, not just limited in search engines, but also in wilder context of automated systems such as recommender systems. For example, Collins et al. \cite{Collins-position-bias} confirmed the position bias in recommender systems, which is the tendency of users to interact with the top-ranked items than the lower-ranked ones, regardless of their relevance. Ovaisi et al. \cite{Ovaisi-selection-bias} focused on the selection bias in the learning-to-rank (LTR) systems, which occurs because ``clicked documents are reflective of what documents have been shown to the user in the first place. '' They proposed a new approach to account of the selection bias, as well as the position bias in LTR systems. Another bias, popularity bias, states the negative influences of historical users' feedback on the qualities of returned items from current recommender systems. Boratto et al. \cite{Boratto-popularity-bias} designed two metrics to quantify such popularity bias and proposed a method to reduce the biased correlation between item relevance and item popularity. 

To reduce biases of the search engines, in other words, is to provide fairer search results. Therefore, our problem is also closely related with fair-ranking studies, in which the goal is to generate ranking lists with nondiscriminatory and fair exposures of various defined groups, such as race, gender, region etc. In our case, the groups are subtopics of the search results and items in each group share similar values and topics. Chen et al. \cite{Chen-gender-bias} investigated the resume search engine and found out the gender-based unfairness from the usage of demographic information in the ranking algorithm. Zehlike et al. \cite{Zehlike-group-fairness} defined the principles of ranked group fairness and the fair top-K ranking problems. They proposed the FA*IR algorithm, which maximizes the utility while satisfying ranked group fairness. In addition to the mitigation of fairness at the group level, Biega et al. \cite{Biega-individual-fairness} proposed new measures to capture, quantify, and mitigate unfairness at the individual subjects level. They proposed a new mechanism --- amortized fairness, to address the position bias in the ranking problems.

Additionally, there are studies in the machine learning community that investigated human's perceptions of fairness and biases in algorithms. Srivastaca et al. \cite{Srivastava-fairness-perception} conducted experiments to detect the most appropriate notions of fairness that best captures human's perception of fairness, given different societal domains. They found out that simplest definition, demographic parity, is aligned with most people's understanding of fairness. Grgić-Hlača et al. \cite{Grgic-Hlaca-fairness-perception} deployed a survey study in the criminal risk prediction domain to analyze how people people perceive and reason the fairness in the decisions generated by algorithms. They found out that people's concerns about fairness are multi-dimensional and unfairness should not be just limited to discrimination.

However, fewer studies in the fair-ranking domain have devoted to probe users' consciousness towards the biases and their behaviors associated with their awareness. Fewer studies have analyzed how users' satisfactions are related with the biases in general. Consequently, inspired by the bias/fairness perception studies in the machine learning community, our work aims to dive deeper in this direction.

\section{Method}\label{method}
In this section, we present the algorithm used to generate the synthesized search pages and the specifics of the survey design. We also enunciate the two core questions we want to address from this study and three hypotheses that we would like to test on. Notice that in sections below, ``diversities'', ``varieties'' and ``differences'' are equivalent. Introducing diversities/varieties/differences could potentially reduce the biases; ``documents'', ``items'' and ``results'' are interchangeable, as they all mean the search engine results.

\subsection*{The Algorithm to Generate Synthesized Pages}

\begin{figure}[ht]
    \centering
    \includegraphics[width=0.9\textwidth]{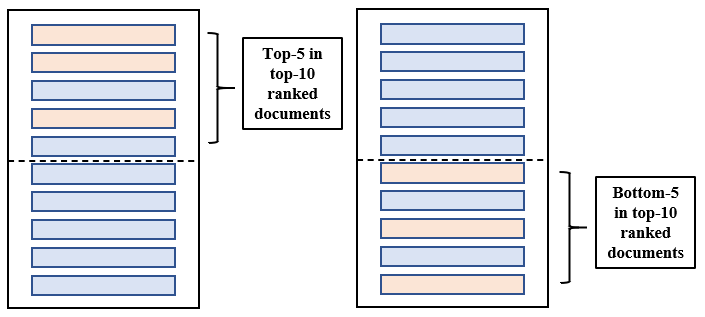}
    \caption{In the left figure, we replace three documents (marked in orange) from top-5 in the top-10 list with lower ranked results. In the right figure, we replace three documents from bottom-5 in the top-10 list.}
    \label{replacement}
\end{figure}

To generate synthesized search pages that are less biased (more diverse in subtopics), we implemented \textbf{epsilon-0.3} algorithm (see Table \ref{algorithm}), with \textbf{statistical parity} as the fairness controller. We first group the documents into $\mathcal{K}$ number of groups. Documents within each group share similar topics, values, views etc. Therefore, each group can be treated as a subtopic group. The fairness controller aims to provide a list of documents with equal or close presence of different subtopic groups: given a search query, we replace three items from the top-10 list with three lower ranked items, proportionally to the frequencies of different subtopic groups in the top-10 list. For instance, suppose that there are two subtopic groups (A and B). If the top-10 list has eight items from group A and two items from group B, we would replace three out of eight items from group A at top-10 with three lower ranked documents from group B. The replacement of the documents could happen in different locations in the top-10 list. Therefore, there are two versions of the algorithm. Version one, presented in Table \ref{algorithm}, replaces three documents from top-5 in the top-10 list. Version two is exactly the same as the version one, except for that the replacement happens at bottom-5 in the top-10 list. Please refer to Figure \ref{replacement} for details.

\begin{table}[ht]
    \footnotesize
    \tabcolsep=0.3cm
    \renewcommand*{\arraystretch}{1.2}
    \centering
    \begin{tabular}{p{12cm}}
        \hline
        \textbf{Epsilon-0.3 Algorithm --- Top-5 Replacement}\\ \hline
        1. Group top-100 search documents into two subtopic clusters.\\
        2. Remove top-10 documents from the two clusters.\\
        3. Calculate cluster 1 (c1) and cluster 2 (c2) frequencies (freq) in top-10 documents.\\
        4. For three iterations:\\ 
        \hspace{0.5cm} \textbf{if} $freq_{c1} == freq_{c2}:$ \\
        \hspace{1cm}Randomly select an index in the top-5 and swap with the next highest ranked \\ \hspace{1cm}document from the cluster, from which the current document was taken. \\ \hspace{1cm}Remove this document from the cluster list \\

        \hspace{0.5cm} \textbf{else if} $freq_{c1} > freq_{c2}:$\\ 
        \hspace{1cm}randomly select an index in the top-5 containing a c1 document, and swap it \\\hspace{1cm}with the next highest ranked document from c2. If there are no remaining \\\hspace{1cm}documents in c2, swap with a c1 document. Remove the swapped document \\\hspace{1cm}from cluster list\\
        
        \hspace{0.5cm} \textbf{else if} $freq_{c1} < freq_{c2}:$\\
        \hspace{1cm}randomly select an index in the top-5 containing a c2 document, and swap it \\\hspace{1cm}with the next highest ranked document from c1. If there are no remaining \\\hspace{1cm}documents in c1, swap with a c2 document. Remove the swapped document \\\hspace{1cm}from cluster list\\
        
        \hspace{0.5cm}Update frequencies \\
        \hspace{0.5cm}Remove the index that has been swapped in the top-5\\ \hline
    \end{tabular}
    \caption{Algorithm to generate synthesized pages. The replacement happens in top-5 from the top-10 list.}
    \label{algorithm}
\end{table}

We chose the \textbf{epsilon-0.3} algorithm not only to be consistent with our previous study, but also based on the fair-ranking work by Gao and Shah \cite{Gao-fair-ranking}. They tested multiple fairness ranking strategies to probe the relationships among fairness, diversity, novelty and relevance and found out that epsilon-greedy algorithms could bring fairer representations of the search results without a cost on the relevance. In the previous study, we experimented with the variants of the algorithm --- epsilon-0.1 and epsilon-0.2, and found out that the replacement ratios (0.1 and 0.2) were too low. Therefore, we decided to work with the epsilon-0.3 algorithm. Additionally, we worked with top-10 list because popular search engines, such as Google and Bing, usually return 10 items per page as the default setting (though adjustable). Therefore, we decided to stick with the default number of 10 documents.

\subsection*{Survey Questions \& Hypotheses}
The two core study questions we would like to answer are: 
\begin{itemize}
    \item \textit{Does introducing more varieties in the search engine results hinder users' satisfaction?}
    \item \textit{What are the reasons of the users' preferences?}
\end{itemize}
We raised three testable hypotheses to potentially answer the questions:
\begin{itemize}
    \item \textbf{H1: People do not care/notice the minute differences between the two search results}: even though we introduced lower ranked results into the top list to add varieties, the differences might not be drastic enough for some participants to notice. Or the participants might realize the differences but they do not care about which one is better.\\
    \item \textbf{H2: The location where the differences present matters. When differences are at the bottom of the search list, people do not care}: intuitively, users might treat the top-ranked results more seriously than the lower-ranked ones. Even in top-10 list, the top-5 might receive different attention than the bottom-5. Therefore, in our survey design, the replacement happens in both locations (top-5 or bottom-5 in the top-10 list). \\
    \item \textbf{H3: People prefer results with high relevance as opposed to high diversity}: this hypothesis could answer the second question. Introducing lower ranked search items means adding more diversities into the results, thus weakening the potential biases of search engines that consistently favor some values over the others. Unavoidably, however, adding lower ranked results would sabotage the relevance of the search results, leading to consequences of potentially lowering users’ satisfactions. Therefore, we want to see whether they prefer higher relevance (more biased) or higher diversity (less biased).
\end{itemize}

\subsection*{Survey Design}
\begin{figure}[ht]
    \centering
    \includegraphics[width=0.8\textwidth]{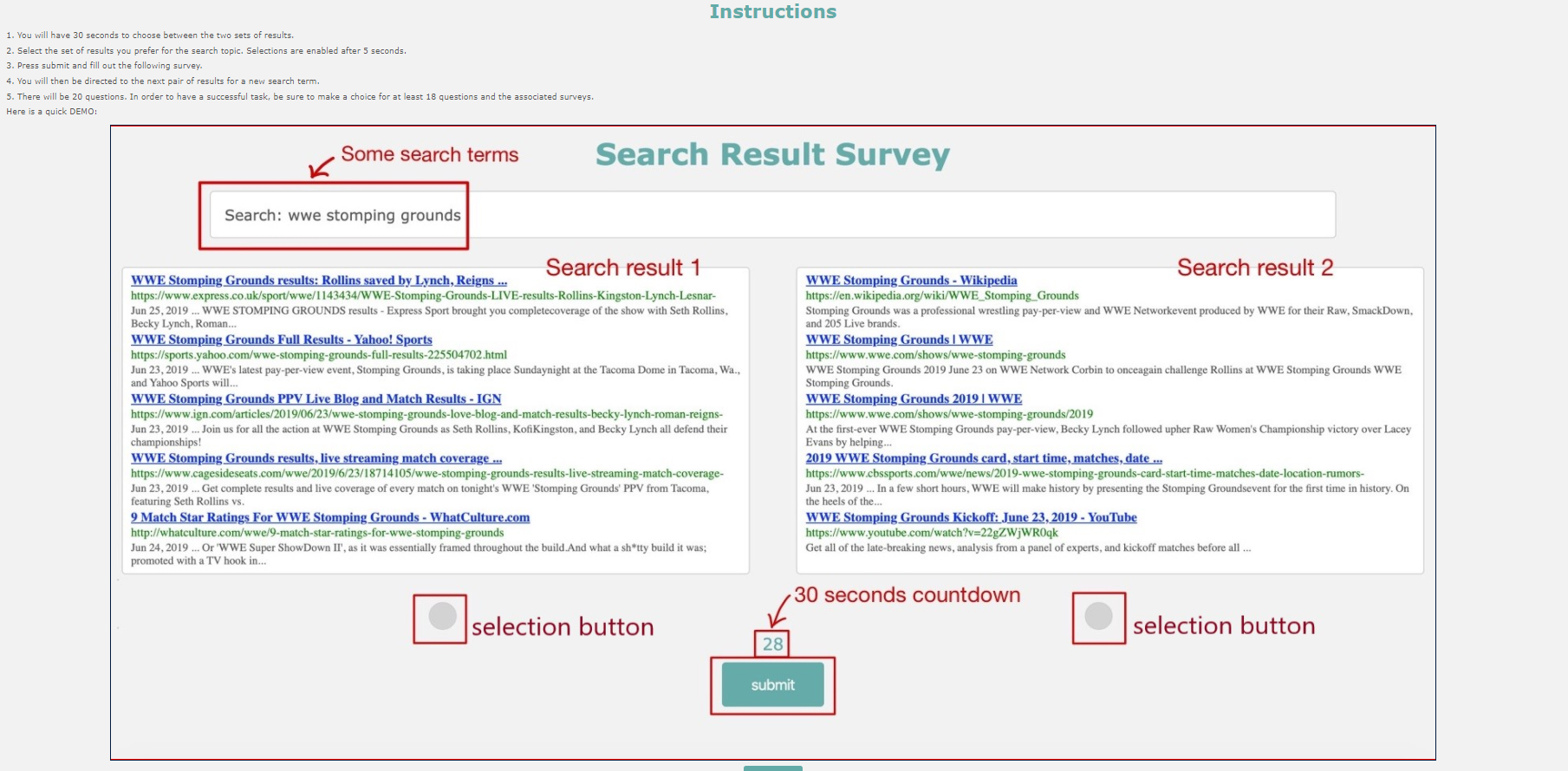}
    \caption{The interface of the survey. The query shows in the search box at the top. The paired pages consist of a real Bing search page and a synthesized one. Participants can select whichever one they prefer and submit.}
    \label{demo}
\end{figure}

The survey starts with a consent form to be signed, which is followed by several demographic questions (age group, gender and education background). Then the participants are provided with instructions on how to complete the survey through a quick demo (as shown in Figure \ref{demo}). Once they are familiar with the details, they may proceed to answer the questions. The survey has 20 rounds in total. Each round consists a pair of real Bing search page and a synthesized page using the algorithm aforementioned, given a specific search query. Participants have 30 seconds to read the query, compare the items between two pages, and make a selection. Out of 20 rounds, we randomly select 10 rounds to perform the top-5 in top-10 replacement, while the rest rounds receive bottom-5 in top-10 replacement.

Based on the experience from the previous study, we thought 20 rounds provide sufficiently large data for statistical analysis, while not fatiguing the participants with too much information. Additionally, with some trial runs, we found out that 30 seconds are enough to participants to compare the two pages and make a selection. After each round, there is a reflection question (as shown in Figure \ref{reflection}) on the reasons of the participants' choice of pages:
\begin{itemize}
    \item \textit{``I did not notice any differences''} addresses H1. The differences might not be palpable enough for the participants. 
    \item \textit{``I noticed some differences but did not care. So I randomly picked up one.''} addresses H1. Participants might detect the discrepancies, but they do not make a difference in users' satisfaction. 
    \item \textit{``I noticed some differences and picked the one that had more results on the same topic.'' \& ``I noticed some differences and picked the one that had more variety in the results.''} They together address H3. More results on the same topic means that the documents are more consistent with each other. More varieties in the results represent the introduced lower ranked results. 
    \begin{figure}[ht]
    \centering
    \includegraphics[width=0.72\textwidth]{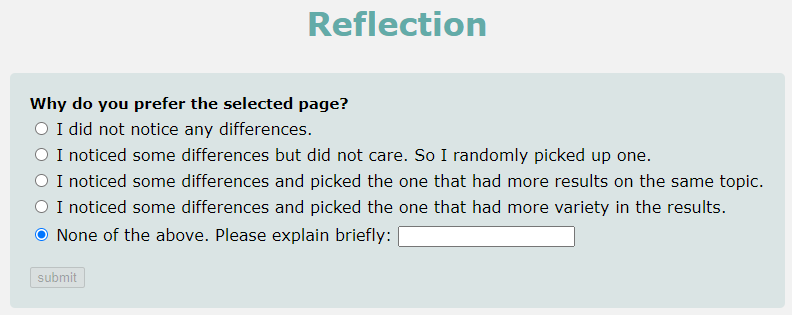}
    \caption{Reflection question on the reasons of the participants' choice of preferences.}
    \label{reflection}
    \end{figure}
\end{itemize}

\section{Results}\label{results}

We launched the survey in MTurk (Amazon Mechanical Turk) by creating a total of 108 assignment. All the assignments were completed within 4 hours from the launch time, with the average completion time as 16.4 minutes. With 137 participants completing the survey, we recorded 2,408 responses. After removing invalid entries, such as users that did not complete the survey or responses with empty selections, 111 participants with 2,134 responses were used in the analysis. The basic demographic information in presented in Table \ref{demographics-table}.

\begin{table}[H]
    \footnotesize
    \tabcolsep=0.4cm
    \renewcommand*{\arraystretch}{1.1}
    \centering
    \begin{tabular}{|c|c|c|}
        \hline
        \textbf{Features} & \textbf{Groups} &\textbf{Count} \\ \hline\hline
        \multirow{5}{*}{\textbf{Age Group}} & 22-28 & 571 (27\%) \\ &29-35 & 681 (32\%) \\ &36-49  & 488 (23\%) \\ &50-65 & 355 (17\%) \\ &Prefer Not To Say & 39 (2\%) \\\hline
        \multirow{4}{*}{\textbf{Gender}} & Male & 1212 (57\%) \\ &Female & 863 (40\%) \\ &Non-binary  & 20 (1\%)\\ &Prefer Not To Say & 39 (2\%)\\\hline
        \multirow{5}{*}{\textbf{Education}} & High School & 158 (7\%) \\ &Bachelor & 1264 (59\%) \\ &Master  & 653 (31\%) \\ &PhD (or similar) & 20 (1\%) \\ &Prefer Not To Say & 39 (2\%) \\\hline
    \end{tabular}
    \caption{Basic demographic information of participants --- age groups, gender, and education background.}
    \label{demographics-table}
\end{table}

\subsection*{Selection Ratios}
Overall, 53.3\% (N=1137) of the responses prefer the real search pages, while 46.7\% (N=997) selected the synthesized versions. We ran Chi-Square goodness of fit test, where the null hypothesis states that the expected frequencies of selecting both choices are the same. The results turned out to be significant at 0.01 significance level (\textbf{p=0.002}). In the bottom-5 replacement group (N=1066), half responses chose the real pages and half chose the synthesized ones. There is no difference in the selection ratios. However, we observed significantly different selection ratios in the top-5 replacement group (N=1068), where 56.6\% (N=604) responses preferred the real pages while 43.4\% (N=464) liked the synthesized pages better. Goodness of fit test yield significant result (\textbf{p}$<$\textbf{1e-4})

Based on the separate tests in each replacement group, it seems that the location where the diversities are introduced have an impact on users' preferences. To further confirm the conjecture, we ran Chi-square test of independence on two categorical variables: users' preferences (real page or synthesized page) and replacement group (top-5 or bottom-5). The result is significant given \textbf{p=0.003}. It demonstrates that the location is associated with participants' preferences.

\subsection*{Reasoning Analysis}
The default four reasons, corresponding to the four choices in order, are ``No Diff'', ``Diff Random'', ``Diff Same Topic'', and ``Diff Variety''. We probed the reasons for three groups separately -- the group that selected the real pages (called ``original'' group), the group that selected the synthesized pages with the top-5 replacement (``Top-5'' group), and the group that selected the synthesized pages with bottom-5 replacement (``Bottom-5'' group). We only presented the analysis of the four default answers here because users' own explanations are diverse, which will be analyzed in the discussion section.

The distributions of default answers for each group are exhibited in Table \ref{reasoning_analysis}. We noticed that within each group, ``Diff Same Topic'' dominated all other answers. Within each group, we ran Chi-square goodness of fit test, in which the null hypothesis states that the expected frequencies of the default choices are the same. All three p-values are extremely small, indicating that the observed frequencies are significantly different from the expected ones. 

\begin{table}[ht]
    \footnotesize
    \tabcolsep=0.03cm
    \renewcommand*{\arraystretch}{1.2}
    \centering
    \begin{tabular}{|c|c|c|c|c|c|}
        \hline
        \textbf{Groups} & \textbf{No Diff} & \textbf{Diff Random} & \textbf{Diff Same Topic} & \textbf{Diff Variety} & \textbf{p-value} \\ \hline
        Original & 117 & 222 & 461 & 310 & 2.2e-49\\\hline
        Top-5 & 49 & 87 & 186 & 132 & 7.5e-20\\\hline
        Bottom-5 & 55 & 104 & 218 & 147 & 1.5e-23\\\hline
    \end{tabular}
    \caption{Distributions of four default choices in each selection group. p-values are from Chi-square goodness of fit test within each group.}
    \label{reasoning_analysis}
\end{table}

\section{Discussion}\label{discussion}

As defined in Section \ref{method}, the first question we would like to answer is \textit{``Does introducing more varieties in the search engine results hinder users' satisfaction?''} From the analysis above, we showed that the proportion of participants preferring the real pages is significantly higher than that of the participants that selected the synthesized pages. Bringing up lower ranked results into the top ranked list introduces more varieties and values in the search results, thus weakening the biases of search engine in favoring certain values. However, it potentially damages the relevance and consistence of the results to the queries. Based on the result, it is reasonable to conjecture that users' satisfactions are impaired due to the added varieties, even though the bias is mitigated. 

The second question we would like to answer is \textit{``What are the reasons of the users' choices of preferences?''}. We hypothesized that the site where the varieties are introduced play a role in affecting users' preferences. And we indeed observed distinctions between the results from the two replacement groups. Our results demonstrated that when the differences exhibit at bottom-5, participants had no compelling tendency of picking up one over the other. However, when they noticed the differences at top-5, they had a significant inclination to choose the real Bing page. Our results are consistent with some previous findings. \cite{Couvering-thesis} articulates that ``most people use very short, imprecise search terms and rarely click beyond the first page of results. They also tend to believe that what comes at the top of the search results must be the best result (p.13).'' Kulshrestha et al. \cite{measure-bias1} also mentioned that the top-ranked results are capable of shaping users' opinions about the topics, which demonstrates the importance of the search results' location. In our case, we could interpret it as when the differences are shown at top-5, participants will pay more attention to pinpoint the differences and make a selection. The phenomenon was also observed in some users' explanations in the reflection questions. Some mentioned that \textit{``I like the first 3 results'', ``\dots at the top that explained \dots, ``preferred the top X list''.}  

Additionally, from the reasoning analysis section, when participants picked up the real pages, 40\% of the reasons are ``Diff Same Topic". It means that they did notice some differences between the two pages and they prefer the one with more homogeneous results. Interestingly, for users that gave their own explanations, many mentioned that the original pages provide more relevant and reliable results than the synthesized ones. Introducing diversities by replacing higher ranked results with lower ranked ones will reduce biases, but potentially hinder the relevance of the search results, thus sabotaging users' satisfaction as the aftermath.

However, when we applied the same reasoning analysis on the two replacement groups, the results do not make logical sense even they are significant. Most of the participants still selected ``Diff Same Topic" as the reason, even though they picked up the synthesized pages that have more varieties rather than consistency. It means they believed that the real pages are more diverse in terms of results. This could be contributed to two reasons: (1) the lower ranked results are similar to the replaced higher ranked items, such that the participants did not notice the diversities; (2) the definitions of similarity and diversity on a topic are not unified and are different from each participant. Consequently, they may pick up the ones that contain subjectively similar results from their perspective, even though the pages are objectively more diverse.

\section{Conclusion}\label{conclusion}
In our study, we designed a survey to assess users' perceptions of search engine biases, with the goal of diagnosing the reasoning underneath their preferences of the real search pages or the synthesized pages. We also investigated the effects of bias-mitigation on users' satisfactions. We noticed that overall, participants prefer the real search pages over the synthesized ones with a significant higher ratio. It indicates that adding more varieties makes the results less biased but less relevant and consistent to the queries, which hurts users' satisfactions. In addition, when the diversities in the synthesized pages are present at the top-5, participants tend to prefer the real pages. However, when they are at bottom-5, there is no significant difference between the ratios of selecting two pages. It confirms our hypothesis as well as some previous studies that the location where the bias-mitigation happens is critical. 

In terms of the future work, two directions could be considered. First, the survey design could be improved. The reflection question in the first round might give additional information of what will be asked in later rounds and could potentially impact users' selections. In addition, the response options in the reflection questions are shown in a fixed order, which might generate order bias \cite{order-bias}. Redesigning the format of reflection questions could potentially improve the study results. Second, if more variables of interests could be collected (if applicable) in addition to the demographic features, mixed-effect regression models could be conducted to account for repeated measures from the same individuals and the relationships among the various features and preferences could be probed simultaneously. 

\bibliographystyle{abbrv}
\bibliography{main.bib}

\begin{thebibliography}{10}

\bibitem{Biega-individual-fairness}
A.~J. Biega, K.~Gummadi, and G.~Weikum.
\newblock Equity of attention: Amortizing individual fairness in rankings.
\newblock {\em The 41st International ACM SIGIR Conference on Research \&
  Development in Information Retrieval}, 2018.

\bibitem{Boratto-popularity-bias}
L.~Boratto, G.~Fenu, and M.~Marras.
\newblock Connecting user and item perspectives in popularity debiasing for
  collaborative recommendation.
\newblock {\em Information Processing \& Management}, 58(1):102387, 2021.

\bibitem{Chen-gender-bias}
L.~Chen, R.~Ma, A.~Hann\'{a}k, and C.~Wilson.
\newblock Investigating the impact of gender on rank in resume search engines.
\newblock In {\em Proceedings of the 2018 CHI Conference on Human Factors in
  Computing Systems}, CHI '18, page 1–14, New York, NY, USA, 2018.
  Association for Computing Machinery.

\bibitem{Collins-position-bias}
A.~Collins, D.~Tkaczyk, A.~Aizawa, and J.~Beel.
\newblock Position bias in recommender systems for digital libraries.
\newblock In G.~Chowdhury, J.~McLeod, V.~Gillet, and P.~Willett, editors, {\em
  Transforming Digital Worlds}, pages 335--344, Cham, 2018. Springer
  International Publishing.

\bibitem{Couvering-thesis}
E.~J.~V. Couvering.
\newblock {\em Search Engine Bias - The Structuration of Traffic on the
  World-Wide Web}.
\newblock {PhD} dissertation, London School of Economics and Political Science,
  2009.

\bibitem{Gao-fair-ranking}
R.~Gao and C.~Shah.
\newblock Toward creating a fairer ranking in search engine results.
\newblock {\em Information Processing and Management}, 57(1), Jan. 2020.
\newblock Publisher Copyright: {\textcopyright} 2019 Elsevier Ltd Copyright:
  Copyright 2019 Elsevier B.V., All rights reserved.

\bibitem{Eric-bias}
E.~Goldman.
\newblock Search engine bias and the demise of search engine utopianism.
\newblock {\em Yale Journal of Law and Technology}, 8, 2005.

\bibitem{Grgic-Hlaca-fairness-perception}
N.~Grgic-Hlaca, E.~M. Redmiles, K.~P. Gummadi, and A.~Weller.
\newblock Human perceptions of fairness in algorithmic decision making: A case
  study of criminal risk prediction.
\newblock In {\em Proceedings of the 2018 World Wide Web Conference}, WWW '18,
  page 903–912, Republic and Canton of Geneva, CHE, 2018. International World
  Wide Web Conferences Steering Committee.

\bibitem{order-bias}
J.~A. KROSNICK and D.~F. ALWIN.
\newblock {AN EVALUATION OF A COGNITIVE THEORY OF RESPONSE-ORDER EFFECTS IN
  SURVEY MEASUREMENT}.
\newblock {\em Public Opinion Quarterly}, 51(2):201--219, 01 1987.

\bibitem{measure-bias1}
J.~Kulshrestha, M.~Eslami, J.~Messias, M.~B. Zafar, S.~Ghosh, K.~P. Gummadi,
  and K.~Karahalios.
\newblock Search bias quantification: investigating political bias in social
  media and web search.
\newblock {\em Information Retrieval Journal}, 22, 2019.

\bibitem{Ovaisi-selection-bias}
Z.~Ovaisi, R.~Ahsan, Y.~Zhang, K.~Vasilaky, and E.~Zheleva.
\newblock Correcting for selection bias in learning-to-rank systems.
\newblock In {\em Proceedings of The Web Conference 2020}, WWW '20, page
  1863–1873, New York, NY, USA, 2020. Association for Computing Machinery.

\bibitem{Srivastava-fairness-perception}
M.~Srivastava, H.~Heidari, and A.~Krause.
\newblock Mathematical notions vs. human perception of fairness: A descriptive
  approach to fairness for machine learning.
\newblock In {\em Proceedings of the 25th ACM SIGKDD International Conference
  on Knowledge Discovery \&amp; Data Mining}, KDD '19, page 2459–2468, New
  York, NY, USA, 2019. Association for Computing Machinery.

\bibitem{sep-ethics-search}
H.~Tavani.
\newblock {Search Engines and Ethics}.
\newblock In E.~N. Zalta, editor, {\em The {Stanford} Encyclopedia of
  Philosophy}. Metaphysics Research Lab, Stanford University, fall 2020
  edition, 2020.

\bibitem{Zehlike-group-fairness}
M.~Zehlike, F.~Bonchi, C.~Castillo, S.~Hajian, M.~Megahed, and R.~Baeza-Yates.
\newblock Fa*ir: A fair top-k ranking algorithm.
\newblock In {\em Proceedings of the 2017 ACM on Conference on Information and
  Knowledge Management}, CIKM '17, page 1569–1578, New York, NY, USA, 2017.
  Association for Computing Machinery.

\end{thebibliography}

\end{document}